\documentclass[aps,pra,amsmath,amssymb,twocolumn,showpacs,nofootinbib]{revtex4} 

\usepackage{graphicx}
\usepackage{dcolumn}
\usepackage{bm}
\usepackage{bbold}
\usepackage{pslatex}
\usepackage{epsfig}
\usepackage{color}
\usepackage{dsfont}

\newcommand{\ket}[1]{\ensuremath{|#1\rangle}}
\newcommand{\bra}[1]{\ensuremath{\langle #1|}}

\newcommand{\be}{\begin{equation}}
\newcommand{\ee}{\end{equation}}
\newcommand{\ba}{\begin{eqnarray}}
\newcommand{\ea}{\end{eqnarray}}

\begin{document}

\preprint{PRE/003}

\title{Environmental correlations and Markovian to non-Markovian transitions in collisional models}

\author{N. K. Bernardes$^{1}$}%
\email{nadjakb@fisica.ufmg.br}

\author{A. R. R. Carvalho$^{2}$}%

\author{C. H. Monken$^{1}$}%

\author{M. F. Santos$^{1}$}%
\email{msantos@fisica.ufmg.br}

\affiliation{$^1$Departamento de F\'isica, Universidade Federal de Minas Gerais, Belo Horizonte, Caixa Postal 702, 30161-970, Brazil}
\affiliation{$^2$Centre for Quantum Computation and Communication Technology, Department of Quantum Sciences, Research School of Physics and Engineering, The Australian National University, Canberra, ACT 0200 Australia }

\date{\today}

\pacs{03.65.Yz 03.67.-a}
\keywords{}

\begin{abstract}
We investigate the smallest set of requirements for inducing non-Markovian dynamics in a collisional model of open quantum systems. This is done by introducing correlations in the state of the environment and analyzing the divisibility of the quantum maps from consecutive time steps. Our model and results serve as a platform for the microscopic study of non-Markovian behavior as well as an example of a simple scenario of non-Markovianity with purely contractive maps, i.e. with no backflow of information between system and environment.
\end{abstract}

\maketitle

\textit{Introduction.} The dynamics of open quantum systems is characterized for continuous Markov processes in terms of master equations in the so-called Lindblad form~\cite{breuer}. However, the required assumption of a memoryless environment is in reality, in most of the cases, an approximation. In general, the environment presents memory effects that may lead to non-Markovian dynamics~\cite{breuer, gardiner, lai, aharonov}. Understanding these effects is an important fundamental question with potential applications in the engineering of reservoirs for quantum computation~\cite{verstraete} and as a resource for quantum information processing such as quantum key distribution~\cite{vasile}, quantum metrology~\cite{matsuzaki,chin}, quantum teleportation~\cite{elsi}, and quantum communication~\cite{Bogna}.

There have been different approaches to study the time evolution of quantum systems subjected to the action of external environments. Some focus on the macroscopic characteristics of the environment, such as its spectral decomposition~\cite{gardiner}, others analyze the general mathematical properties of the quantum maps they produce~\cite{gorini, lindblad, wolf, wolf2}. In a line of possible approaches, one could argue that these two form the borders, each of which presenting powerful but yet incomplete pictures. The mathematical approach, based on the infinite divisibility of the time evolution into trace-preserving completely positive quantum maps (from now on CP maps), is formally absolute but difficult to connect to practical examples. On the other hand, the macroscopic approach relates to many experiments realized (or realizable) in labs but can only draw generic pictures, failing to address the detailed microscopic origins for the behavior of the system. 

\begin{figure}[t]
	\centering
		\includegraphics[width=0.35\textwidth]{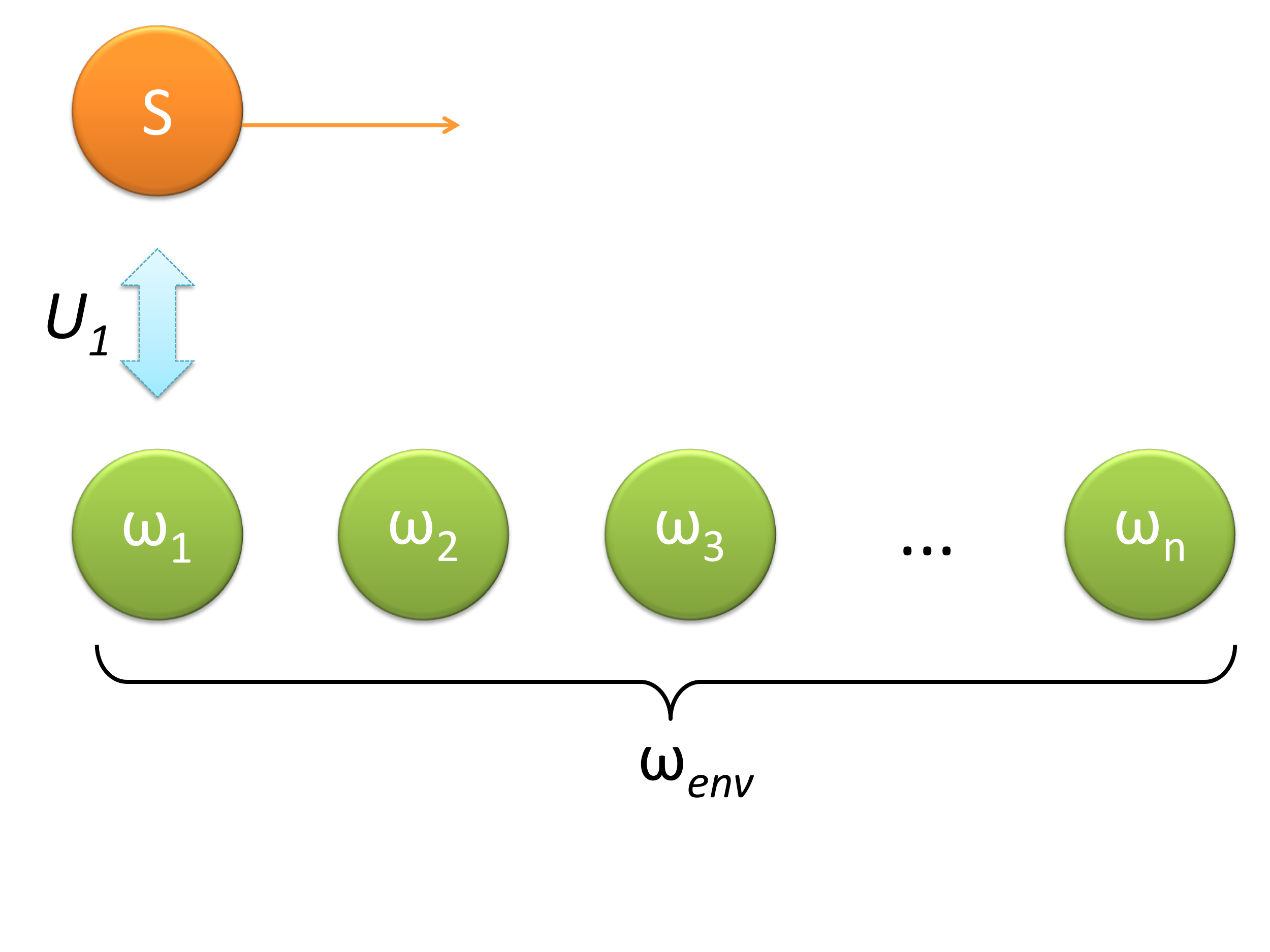}
	\caption{(Color online) Schematic picture of the collisional model.}
	\label{scheme}
\end{figure}

There is a third approach, based on collisional models, {\color{black} which initially was proposed to study the relaxation phenomena of simple systems as coupled spins and coupled harmonic oscillators~\cite{rau}} and has been used in Cavity QED experiments for decades as an effective way to simulate Markovian reservoirs~\cite{sargent, brune, haroche1, haroche2, haroche3}. In most of these models, the system $\rho$ is made to interact (collide), one at a time and sequentially, with particles $\omega_1, \omega_2...\omega_n$ that form a controlled environment. Each of these interactions of strength $\eta_i$ and duration $\tau_i$ is described by a unitary transformation $U_i$, as illustrated in Fig.~\ref{scheme}. It is assumed that each environmental particle interacts only once with the system and that initially system $\rho$ and environment $\omega_{env}$ are decoupled, $\rho\otimes \omega_{env}$. Under these assumptions the evolution of the system after the j$th$ collision is given by
\be\label{rhof}
\rho\rightarrow\rho'=\text{Tr}_{env}\left[U_j...U_1(\rho\otimes \omega_{env})U_1^{\dagger}...U_j^{\dagger}\right],
\ee
where $\text{Tr}_{env}$ denotes the partial trace over the environmental degrees of freedom. In standard collisional models the environmental particles are in a factorized state $\omega_{env}=\omega^{\otimes n}$ and in the limit of vanishing product $\eta \tau$, the effective evolution of the system can be approximated by a standard Lindblad form~\cite{ziman1, ziman2}.

More recently, collisional models have also drawn theoretical attention for their potential advantages in the microscopic study of environmental memory effects. In particular, many recent theoretical works have connected variations of collisional models to the study of non-Markovianity~\cite{giovannetti, tomas, ciccarello, vacchini, budini, Paternostro} exploring the specific roles played by the reservoir in the dynamics of the system. There are many different ways in which a reservoir may induce a non-Markovian time evolution on a given quantum system, correlations in its quantum state, time dependent interactions with the system, and the internal dynamics of the reservoir itself being the most common~\cite{breuer, gardiner}. {\color{black} In most of the papers studying these models, the authors either focus on building up and understanding a memory kernel function (MKF)~\cite{giovannetti,ciccarello, vacchini, budini} or they study non-Markovian effects that arrive from the accumulation over many collisions and that are due to strong long range correlations in the environment~\cite{tomas,Paternostro}. The first one involves studying some integro-differential type of Master equation and deriving general conditions on the MKF present in this equation so that the evolution of the system is either Markovian or non-Markovian. This approach can be connected to standard quantum statistical general properties such as the effects of sub or super-ohmic reservoirs and so on. The second approach, on the other hand, produces an effect equivalent to the one obtained when the system interacts with a low dimensional environment (a qubit in the limiting case) which is as far from a thermodynamical reservoir as possible.

However, and more important for us, it happens that in both cases the three main ingredients that may lead to a non-Markovian evolution, namely the correlations in the environment, its internal dynamics and the detailed interaction between the system and the environment, they all end up bunched together as a sole effect. For example, these effects combine to form the kernel function of the MKF based results, and from then on the analysis focus solely on its general properties. On the other hand, in the Ólong range correlationÓ works, they also combine, this time to produce an overall periodic system-environment dynamics that cannot be described by a Markovian evolution of the system alone (see, for example, Ref.~\cite{tomas}).

In this paper, we would like to address a different set of questions: keeping the system-reservoir interaction constant and assuming no internal dynamics of the reservoir between collisions, which would be the minimal amount of correlation in the environmental state in order to generate a non-Markovian evolution of the system? Would any correlation be sufficient? Which type of correlation? How would it depend on the specific details of the interaction between system and environment? Which would be the order of magnitude of these non-Markovian effects? 

In order to answer these questions, we divide our paper as follows: first, we design a new collisional model that is able to generate, in principle, any unital open system dynamics on a target qubit and, in particular, a master equation in the Lindblad form. Then, we introduce small modifications to the model, specifically concerning correlations in the environmental state to study the influence of these modifications in the emergence of memory effects and on the eventual induction of non-Markovian evolutions in the system. In order to do so, we adopt the general criterion of divisibility of quantum maps, defined in~\cite{wolf}, used as a measure in~\cite{rivas} and extensively reviewed in section 3 of Ref.~\cite{Plenio3}, and relate it to the properties of the Choi matrix (also known as the dynamical matrix)~\cite{choi, bengtsson, simon}. Finally, we analyze the evolution of the correlation between the system and the environment and whether there is backflow of information to the system in the non-Markovian regime.} 

{\color{black}Note that the relation between the system-environment correlation and non-Markovianity has been investigated in Refs.~\cite{modi, rodriguez-rosario}. However, these works do not focus on how the details of the internal properties of the environment and its interaction with the system will influence the dynamics of the latter which are exactly the answers we seek in this work.}


\textit{Model.} {\color{black} The first goal is to show that our model reproduces the master equation in its most standard form and that all the three ingredients are easily and independently modifiable.} For simplicity and experimental reproducibility, we take one qubit ($\{|0\rangle,|1\rangle\}$) as our system and consider its sequential interaction, one at a time, with an environment made of a string of qudits ($\{|0\rangle,|1\rangle,|2\rangle,...,|d-1\rangle \}^{\otimes n}$, $n\rightarrow \infty$). The system is initially decoupled from the environment whose state before the interactions is given by $\omega_{env} = \omega^{\otimes n}$ where {\color{black} $\omega = (1-\sum_i \epsilon_i)\ket{0}\bra{0}+\sum_i \epsilon_i\ket{i}\bra{i}$, $0\leq\epsilon_i \leq 1$, $0\leq \sum_i \epsilon_i \leq 1$} and $i=1..d-1$. The interaction Hamiltonian for each collision is given by $H=\eta(\mathds{1}\otimes\ket{0}_R\bra{0}+\sum_i \sigma_i\otimes\ket{i}_R\bra{i})$, in units of $\hbar$. The $\sigma_i$ matrices are unitary operators that act on the system and ``point'', for now, into arbitrary directions given by unitary vectors $\hat{s}_i$, that is, $\sigma_i = \vec{\sigma}.\hat{s}_i$, where $\vec{\sigma}=(\sigma_x,\sigma_y,\sigma_z)$. {\color{black} Note that $H^2 \propto \mathds{1}_{s}\otimes \mathds{1}_{R}$, therefore, the overall time evolution operator $U$ for any given collision can be expanded as 
\begin{equation}
U(\tau)= \cos (\eta\tau) - i \frac{H}{\eta} \sin (\eta\tau).
\end{equation}
Each collision is then set to take a time $\tau = \pi / (2 \eta)$ so that the state of the system after a collision is given by 
\ba
\rho(t+\textrm{collision})=\text{Tr}_{env}\left[U(\tau)(\rho(t) \otimes \omega)U(\tau)^{\dagger}\right]\nonumber\\
=(1-\sum_i \epsilon_i)\rho(t)+\sum_i \epsilon_i\sigma_i\rho(t)\sigma_i.
\label{rho(1)}
\ea
The differential standard master equation in the Lindblad form for different unital channels can be written as $\rho(t+\Delta t) = \rho(t) - \sum_i \gamma_i \Delta t \rho(t) + \sum_i \gamma_i \Delta t \sigma_i \rho(t)\sigma_i$ where $\gamma_i \Delta t$ is a dimensionless quantity related to the probability of channel ``\textit{i}'' changing the state of the system in time $\Delta t$, the relative rate of change between different channels being given by $\gamma_i/\gamma_j$ or, equivalently by $\gamma_i \Delta t/(\gamma_j \Delta t)$. A quick look in the equation produced by our model shows that a simple identification of $\epsilon_i \equiv \gamma_i \Delta t$ provides exactly the same evolution for the system after any given collision so that the state of the system after the $j$th collision reads,
\ba
\frac{\rho(T+\Delta t)-\rho(T)}{\Delta t}= 
-(\sum_i \gamma_i)\rho(T)+\sum_i \gamma_i \sigma_i \rho(T)\sigma_i,
\label{Lindblad1}
\ea
where $T=(j-1)\Delta t$ and $\epsilon_i \equiv \gamma_i \Delta t$. This is physically reasonable: $\epsilon_i$ is encoded in the environmental state and gives exactly the probability of channel ``\textit{i}'' acting on the system. Besides, in the limit of $\Delta t \rightarrow 0$ ($\epsilon_i\rightarrow 0$), this is the differential version of a standard master equation corresponding, therefore, to a single time step physical implementation of a CP map. Note that because $\epsilon$ can be as small as needed, this environment, albeit highly organized, still truthfully reproduces the effects and the physics of $d$ independent unital ($\sigma_i^\dagger\sigma_i=\textit{I}$) Markovian reservoirs where the overall time interval $T$ is determined by the number of collisions. Also note that, contrary to the standard collisional models, as in Refs.~\cite{giovannetti, tomas, ciccarello, vacchini, budini,Paternostro}, we rely not on the duration of the collision itself (which is always the same) but on the state of the reservoir in order to map the evolution of the system onto the standard Lindblad form{\color{black}\footnote {\color{black} {The main effect of different choices of $\tau$ would be to add terms of the form $\epsilon_i[\sigma_i, \rho(t)]$ to Eq.~(\ref{rho(1)}). These terms, that could be seen as Hamiltonian terms generated by the collision, would not change the results presented in this work but explicit expressions for state $\rho(T+2\Delta t)$ would become much more complicated to analyze and deviations from the standard master equation due to non-Markovian effects much more difficult to understand.}}}. Finally, note that taking the same state $\omega$ for each environmental particle is equivalent to choosing constant decoherence rates for the evolution of the system.

There are some clear advantages for using this model to simulate reservoirs and specially to study Markovian to non-Markovian transitions. First of all, it connects directly even to the more restrictive mathematical description of Markovianity (as described above) and is simple enough to implement in basically any quantum optical setup {\color{black}(same concepts used in~\cite{Liu, Chiuri, Steve}, for example)}. Second, it is flexible enough to allow for the generation of basically any unital time evolution of the system: through a suitable choice of the interaction Hamiltonian and the specific set of $\epsilon_i$ encoded in $\omega$ for each environmental particle, one can adjust the number and type ($\sigma_i$) of independent channels acting on the system as well as their effective $\gamma_i$ rates as a function of time. {\color{black} And, because time intervals can be chosen as small as necessary ($\Delta t \propto \epsilon$) and the effective number, type and strength of the channels can be controlled, step by step, by choosing the respective $\epsilon_i$ to be either finite or zero, one can use our model to design any unital map from one collision to the following which translates into any unital overall open system dynamics for the target qubit.}  Third, note that we can also take the reservoir particles to be in a pure state, making it easy to calculate correlations among them and also between the system and the reservoir after each collision. For example, state $\omega$ or its pure state version $\omega'=\ket{R}\bra{R}$, where$\ket{R}=\sqrt{1-\sum_i \epsilon_i}\ \ket{0}+\sqrt{\epsilon_i}\ \ket{i}$, produce exactly the same $\rho(\Delta t)$ after the first collision. Finally, also note that, in its mixed state version, the environmental state $\omega^{\otimes n}$ does not change in time, another desirable characteristic of a true reservoir. 

\textit{Non-Markovianity.} In its standard version, presented above, our model renders the Lindblad form. However, and more importantly, it can be easily modified to generate memory effects that may ultimately lead to non-Markovian evolutions of the system. One possibility is to increase $\epsilon_i$ (change state $|R\rangle$) of at least one environmental particle in such a way as to not obtain the differential form of the master equation after the corresponding collision. This, however, creates a somewhat \textit{trivial} non-Markovianity that arises even for a ``single particle'' environment which is as far as conceivable from a true reservoir. This is the origin of the non-Markovian effect in~\cite{Steve, Paternostro} for example. On the contrary, we want to understand memory effects due to correlations in the environment while preserving its reservoir-like characteristics. In particular, the one that guarantees small changes to the system in each time step which translates to $\epsilon_i \ll 1$ in our model.

Even within these restrictions, there are still many different ways to correlate the state of the environment, each leading to different dynamics of the system. {\color{black} For instance, one possibility similar to the case studied in~\cite{tomas}, is to collide the system with infinitely correlated particles which, in our model, would translate into an environmental state $\omega_{env}= \sum_i p_{i} |ii...iiii..\rangle\langle ii...iiii..|$ where $i=0,1,2,...d-1$. Note, however, that the effect of such environment after $n$ collisions is to generate state $\rho(t+n\Delta t) = p_0 \rho(t) + \sum_{i=1}^{d-1} p_i \sigma_i^n \rho(t) \sigma_i^n$ for the system which is either $\rho(t)$ for $n$ even (because $\sigma_i^2 = 1$) or the single collision state $\rho(t+n\Delta t) = p_0 \rho(t) + \sum_{i=1}^{d-1} p_i \sigma_i \rho(t) \sigma_i$ if $n$ is odd. This dynamics would be naturally non-Markovian but in the same trivial way of the ``single particle, large $\epsilon$'' scenario analyzed in the previous paragraph. 

Here, we are interested in identifying the smallest set of conditions for non-Markovian behavior in our model, therefore we will ask the simplest question}: when can correlations between two consecutive collisions generate memory effects? In order to answer this question, we will analyze, from now on, the two-channel environment made of a sequence of qutrits, for reasons that become clear later. In this scenario, two correlated consecutive collisions are obtained if one replaces any two neighboring $\omega \otimes \omega$ portion of the environment by a two-particle state $\Omega = \ket{R_2}\bra{R_2}$ where 
\begin{align} \label{R2}
&\ket{R_2}=(1-\epsilon_1-\epsilon_2)\ket{00} + \sqrt{1-\epsilon_1-\epsilon_2}(\sqrt{\epsilon_1}\ket{01}+\sqrt{\epsilon_2}\ket{02}) \nonumber\\
&+ \sqrt{\epsilon_1}\sqrt{1-2[q\epsilon_1+(1-q)\epsilon_2]}\ket{10} \\
&+ \sqrt{\epsilon_2}\sqrt{1-2[q\epsilon_2+(1-q)\epsilon_1]}\ket{20} \nonumber \\
&+ \sqrt{2(1-q)\epsilon_1\epsilon_2}(\ket{12}+\ket{21}) +\sqrt{2q}(\epsilon_1\ket{11}+\epsilon_2 \ket{22}), \nonumber
\end{align}
and $0 \leq q \leq1$, as shown in Fig.~(\ref{scheme2}). A quick inspection of this state shows that when $q=1/2$ it reduces to the uncorrelated product $|R\rangle |R\rangle$ and when $q>1/2$ ($q<1/2$) the collisions are correlated (anti-correlated). 
Also note that for any $q$, state $\rho(T+\Delta t)$ still follows Eq.~(\ref{Lindblad1}) for the chosen $\sigma_{1,2}$ channels and, as expected,  correlation will affect the system only after the second collision. Finally, note that $\Omega$ or its mixed version made of just its diagonal terms in the same $\{|00\rangle,|01\rangle,...\}$ basis produce, once again, exactly the same dynamics on the qubit system. 
\begin{figure}[t]
	\centering
		\includegraphics[width=0.35\textwidth]{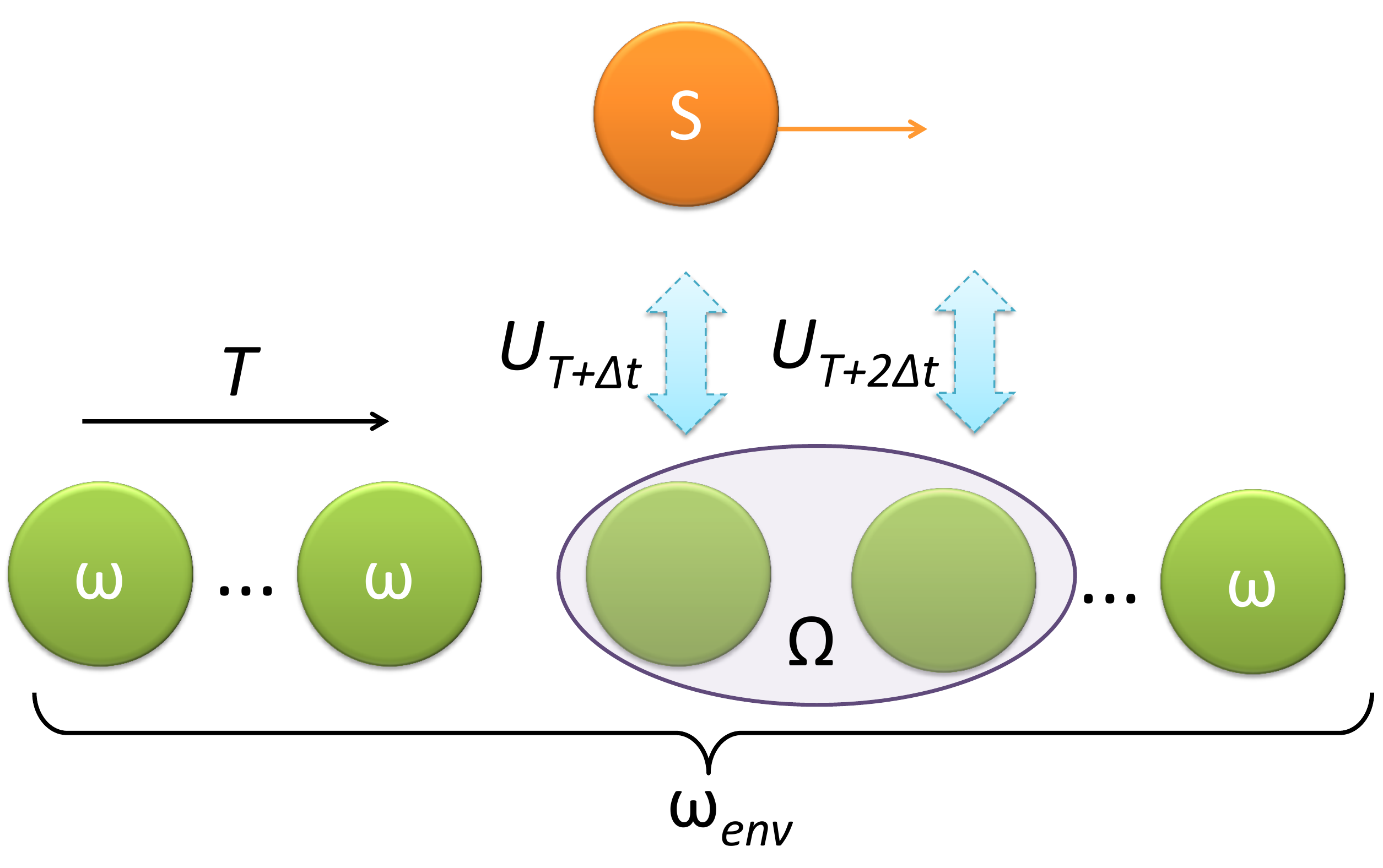}
	\caption{(Color online) Schematic picture of our collisional model with an environment that is correlated between $T$ and $T+2\Delta t$}
	\label{scheme2}
\end{figure}

On the other hand, when $q \neq 1/2$, there are three independent sets of parameters that can influence the dynamics of the system: $\{\epsilon_1,\epsilon_2\}$, $\{\sigma_1,\sigma_2\}$ and $q$ each translating respectively into different decoherence rates, different unitaries induced on the system and the quantity of correlation in the environmental state. Similarly to Eq.~(\ref{rho(1)}) and provided $|R_2\rangle$ is used, the state of the system after two time steps reads:{\color{black}
\begin{align}\label{rho3}
&\rho(T+2\Delta t)=\textrm{Tr}_{env}\left[U(\tau)U(\tau)(\rho(T)\otimes\ket{R_2}\bra{R_2})U^{\dagger}(\tau)U^{\dagger}(\tau)\right]\nonumber\\
&=\left(1-\epsilon_1-\epsilon_2\right)\left[(1-\epsilon_1-\epsilon_2)\rho(T)+\epsilon_1 \sigma_1\rho(T)\sigma_1\right.\nonumber\\
&\left.+\epsilon_2 \sigma_2\rho(T)\sigma_2\right]+2q(\epsilon_1^2+\epsilon_2^2)\rho(T)+\\
&+ \epsilon_1\left\{1-2[q\epsilon_1+(1-q)\epsilon_2]\right\}\sigma_1\rho(T)\sigma_1\nonumber \\
&+ \epsilon_2\left\{1-2[q\epsilon_2+(1-q)\epsilon_1]\right\}\sigma_2\rho(T)\sigma_2 \nonumber \\
&+ 2(1-q)\epsilon_1\epsilon_2(\sigma_1\sigma_2 \rho(T) \sigma_2\sigma_1 + \sigma_2\sigma_1 \rho(T) \sigma_1\sigma_2), \nonumber
\end{align}
where $\rho(T)$ is the state of the system immediately before interacting with the correlated part of the environment. We can rewrite this state in terms of $\rho(T+\Delta t)$ as follows
\begin{align}\label{rho3}
&\rho(T+2\Delta t)=\rho(T+\Delta t) + \epsilon_1 \mathcal{L}_1 \rho(T+ \Delta t) + \epsilon_2 \mathcal{L}_2 \rho(T+\Delta t)\nonumber\\
&+(2q-1) \{(\epsilon_2-\epsilon_1)[(\epsilon_1\mathcal{L}_1\rho(T)-\epsilon_2\mathcal{L}_2\rho(T)] \nonumber\\
&+\epsilon_1\epsilon_2[2\rho(T)-\sigma_1\sigma_2 \rho(T) \sigma_2\sigma_1 - \sigma_2\sigma_1 \rho(T) \sigma_1\sigma_2]\},
\end{align}
where $\mathcal{L}_i \rho = -\rho + \sigma_i \rho \sigma_i$.} First thing to notice is that, as expected, when $q=1/2$ we recover the original Lindblad form for the second time step, i.e. any non-Markovian behavior must come from the terms proportional to the correlation factor $Q \equiv (2q-1)$. Also note that these terms are of the order of $\epsilon^2$ and, therefore, are much smaller than the terms of the Markovian time evolution obtained for $q=1/2$ ($Q=0$). As we analyze later on, that means that there is no backflow of information from the reservoir to the system in consecutive time steps and any non-Markovianity will not be witnessed by measurements such as the distinguishability of two arbitrary states~\cite{breuer2}. This type of non-Markovianity has been analyzed recently in~\cite{sabrina} and our model provides a simple framework to study its consequences and characteristics. 

One can use Eq.~(\ref{rho3}) to calculate $\rho(T+2\Delta t)$ for an arbitrary $\rho(T)$ and, with the corresponding $\rho(T+\Delta t)$, extract the overall map $\Phi_{21}$ that implements $\rho(T+2\Delta t)=\Phi_{21}[\rho(T+\Delta t)]$. Testing Markovianity of these two collisions translates into checking the complete positivity of $\Phi_{21}$. We discuss how to do this and the most general maps later on. But first, we will simplify the problem to two different scenarios that already illustrate all the important effects brought by the correlation terms in the dynamics of the system. And, because $\rho(T)$ is arbitrary, we will set $T=0$ from now on with no loss of generality. 

\textit{Same channels.} The first case is obtained when the two standard channels are the same, i.e. $\sigma_1=\sigma_2=\sigma$. In this case, correlation can only arise for different parameters $\epsilon_1 \neq \epsilon_2$ such that if the system changes by a certain amount in the first collision it is more (or less) likely to change by that same amount in the second collision. Under these conditions, Eq.~(\ref{rho3}) can be rewritten as  
\ba
\frac{\rho(2\Delta t)-\rho(\Delta t)}{\Delta t}=\gamma_q[-\rho(\Delta t)+\sigma\rho(\Delta t)\sigma], \label{e1-e2}
\ea
where $\gamma_q=\gamma_1+\gamma_2-Q(\gamma_1-\gamma_2)(\epsilon_1-\epsilon_2)/[1-2(\epsilon_1+\epsilon_2)]$. This equation is still in the Lindblad form but with a modified rate which means that, for channels of the same type, the effect of $Q \neq 0$ ($q \neq 1/2$) in consecutive collisions is just to modulate the decoherence rate while still preserving the overall Markovianity of the evolution. Also note that this correction is of the order of $\gamma \epsilon$ ($\epsilon \ll 1$), i.e. it is much smaller than the effect of the standard, uncorrelated, reservoir. 

\textit{Same coefficients.} The other simple and meaningful scenario happens when the two decoherence rates of the standard channels are the same, $\epsilon_1=\epsilon_2=\epsilon$ in which case correlation necessarily means different channels acting on the system ($[\sigma_1,\sigma_2] \neq 0$). At this point, just for the sake of simplicity but with no loss of generality, we will set $\sigma_1 = \sqrt{a}\ \sigma_x + \sqrt{1-a}\ \sigma_z$ and $\sigma_2=\sigma_z$. Note that these two channels correspond to axes in the Bloch sphere and any arbitrary non-commuting set $\left\{\sigma_1,\sigma_2\right\}$ can always be rotated into our particular choice. The general solution in this case is presented in the Appendix but, more importantly, under our primary condition of $\epsilon \ll 1$, it reads
\ba
\frac{\rho(2\Delta t)-\rho(\Delta t)}{\Delta t}= \gamma(\mathcal{L}_1+\mathcal{L}_2)\rho(\Delta t) \nonumber \\
-2\gamma a\epsilon Q \left[-\rho(\Delta t)+\sigma_y\rho(\Delta t)\sigma_y\right],
\label{rhoNM}
\ea 
where we keep terms up to $O(\epsilon^2)$. Some things pop out immediately when looking at Eq.~(\ref{rhoNM}): first, and as expected, when $Q=0$ (or $a=0$), it reduces back to the Markovian evolution of two independent channels (or one). Second, the term proportional to the correlation factor $Q$ is, again, of the order of $\gamma \epsilon$ and, hence, much smaller than the ones generated by the uncorrelated evolution. Third, and more important, this term looks like an effective reservoir created by the correlation but it is only in the Lindblad form when $Q<0$ ($q<1/2$), i.e. when the environment is anti-correlated. When $Q>0$ ($q>1/2$), the effective ``decoherence'' rate $\gamma_{\mbox{\scriptsize{eff}}} = -2\gamma a\epsilon Q$ is negative and the map $\Phi_{21}$ is not CP, i.e. the evolution is non-Markovian. Note that the non-Markovian transition exists for any value of $a \neq 0$ (i.e. for any non-commuting set ${\sigma_1,\sigma_2}$), whose effect is just to modulate the $\gamma_{\mbox{\scriptsize{eff}}}$ created by the correlation. Also note that the addition of an extra Markovian channel in the $y$ direction, $\mathcal{L}_y$, of rate $\gamma_y$, would require larger values of $Q$ for the non-Markovian evolution to happen. In this case, the evolution would be non-Markovian only for $Q > \gamma_y/(2a\gamma\epsilon)$ and would be strictly Markovian as soon as $\gamma_y = 2a\gamma\epsilon$. In our model, that would correspond to adding an extra level to the environmental particles (ququarts instead of qutrits), making $\sigma_3 = \sigma_y$, $\epsilon_3 =\gamma_y \Delta t$ and adjusting state $|R_2\rangle$ to include this extra channel. Finally, note that the fact that it is correlation and not anti-correlation that generates non-Markovianity is a consequence of the particular choices we took in our model, such as the specific interaction Hamiltonian $H$. What is important is the transition from one regime to the other depending on the parameters that dictate the time evolution of the system and that are contained not only in $H$ but specially in $\omega_{env}$.

These two scenarios contain all the effects that correlations in two consecutive collisions may generate in the evolution of the system: modulations of the standard decoherence rates and/or the creation of effective channels in the directions orthogonal to the standard ones. Both effects ($O(\epsilon^2)$) are much smaller than the standard changes to the state of the system (of the order of $\epsilon$), which means that only the second one can be associated to a non-Markovian evolution -- the first effect is equivalent to just introducing a small change in the already existing $\epsilon_i$ associated to the second collision. Here, we considered only two standard channels, $\sigma_{1,2}$, but the analysis in this paragraph is general and it holds for an arbitrary number of channels in the model: correlations between any two of them will be equivalent to redefining their standard decoherence rates relative to the second collision and to creating an effective channel in an orthogonal direction. This effective channel may not represent a CP map, i.e. it may feature a negative decoherence rate and that will translate into the non-Markovian evolution of the system as long as its modulus is larger than an eventual standard rate in the same direction. The example that leads to Eq.~(\ref{rhoNM}) features the most favorable case for non-Markovianity in our model, where the standard rate of the orthogonal channel is actually equal to zero, i.e., there is no standard $\sigma_y$ channel to compete with the non-Markovian effect brought by correlations in the environmental state.  

These simple cases already encompass the important properties of the model. However, one can still solve the most complete scenario by extracting the general map $\Phi_{21}$ from Eq.~(\ref{rho3}) and $\rho(\Delta t)$. By construction, our model produces a CP map $\Phi_{20}$ that takes state $\rho(0)$ into Eq.~(\ref{rho3}) and based on the divisibility criterion~\cite{wolf, Plenio3}, this map is Markovian if it is divisible into two consecutive CP maps $\Phi_{20}=\Phi_{21}\Phi_{10}$. $\Phi_{10}$ is CP by construction and the question reduces to the characteristics of $\Phi_{21}$. The density matrix of a qubit state can be represented by $\rho=\left(\mathbb{1}+\vec{r}\cdot\vec{\sigma}\right)/2$, where $\vec{r}$ is the Bloch vector ($r_i=\textrm{Tr}(\rho \sigma_i)$). The action of a map $\Phi$ on a qubit quantum state will be given, in general, by $\Phi:\vec{r}\mapsto\vec{r'}=\Lambda\vec{r}+\vec{t}$, where $\Lambda$ is responsible to change the norm and rotate the Bloch vector and $\vec{t}=(t_x,t_y,t_z)$ to translate it. For unital maps ($\vec{t}=0$), we can associate to this map an Hermitian matrix $\mathcal{H}=\left(\mathds{1}+\sum_{\mu,\nu=x,y,z}\Lambda_{\mu\nu}\sigma_{\mu}\otimes\sigma^{*}_{\nu}\right)/2$. The map $\Phi$ is completely positive iff $\mathcal{H}\geq0$ \cite{choi, simon}. We solved this problem in our model by extracting the matrix $\mathcal{H}$ from our map and diagonalizing it just to find that only one of its eigenvalues can be smaller than zero. In Fig.~(\ref{figineq2}) we plot this eigenvalue $\lambda$ as a function of $Q$ and $a$ and also as a function of $Q$ for one particular value of $a$, for different $\epsilon_1,\epsilon_2$. As already anticipated, $Q=0$ ($q=1/2$) draws the line of non-Markovian to Markovian transition and any $a\neq 0$, corresponding to non-commuting uncorrelated channels, only defines the intensity of the effective reservoir. 
\begin{figure}[t!]
	\centering
	\includegraphics[width=0.35\textwidth]{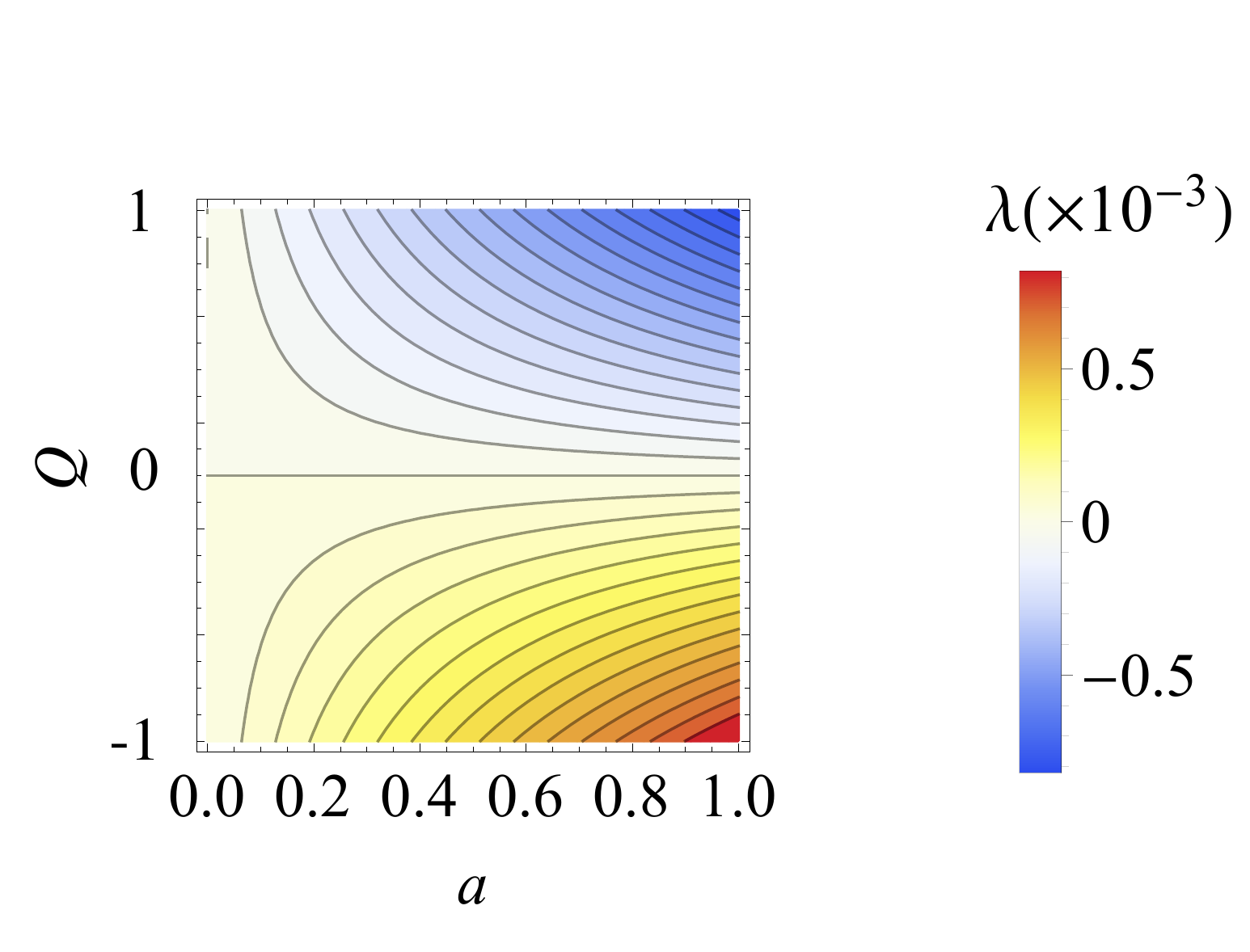} \includegraphics[width=0.35\textwidth]{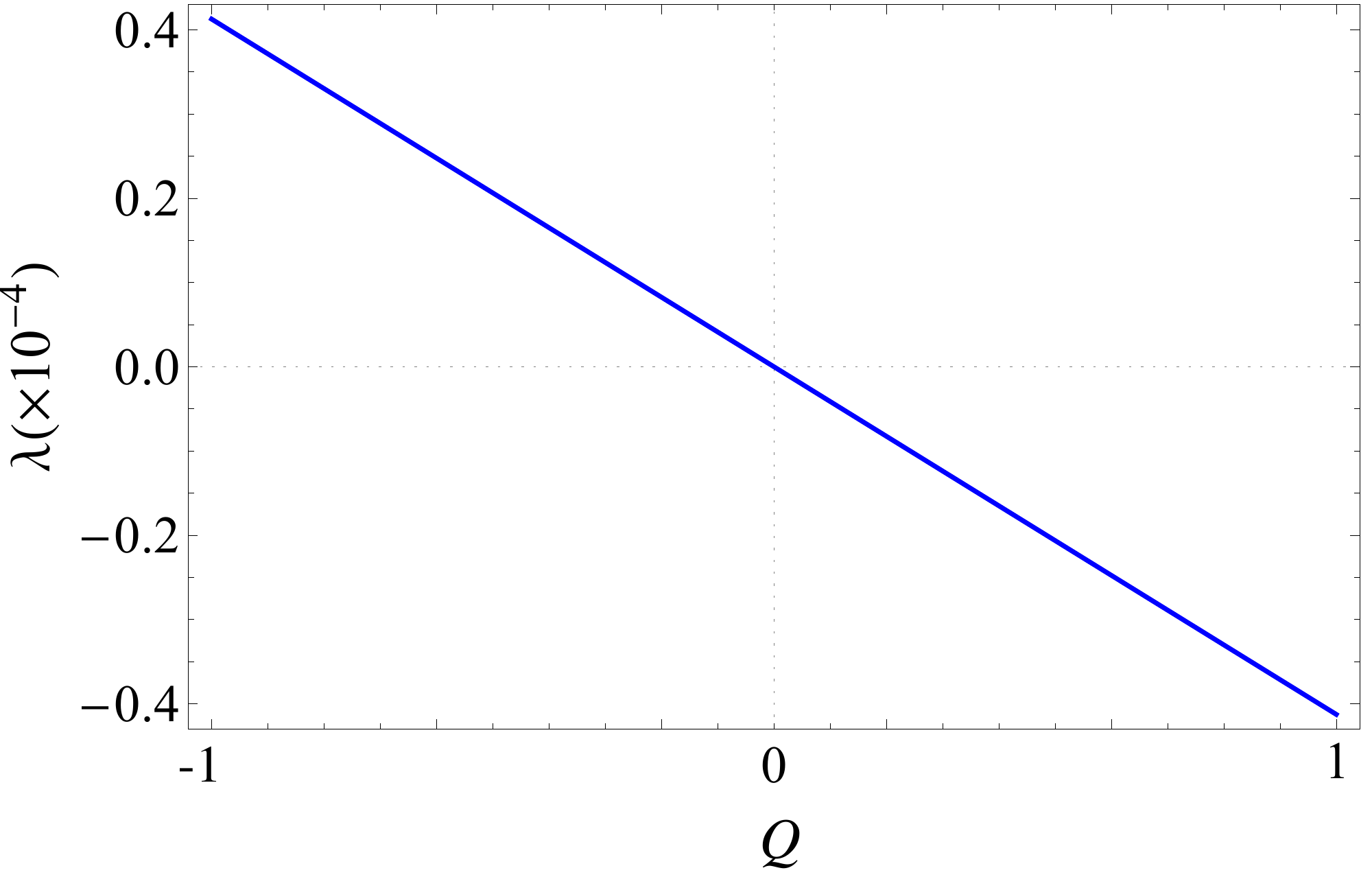}
	\caption{(Color online) Eigenvalue of $\mathcal{H}$ in terms of the correlation factor $Q$ and parameter $a$ for $\epsilon_1=0.01$ and $\epsilon_2=0.02$ (top). Eigenvalue of $\mathcal{H}$ in terms of the correlation factor $Q$ for $a=0.05$, $\epsilon_1=0.01$, and $\epsilon_2=0.02$ (bottom).}
	\label{figineq2}
\end{figure}
Finally note that even though we are particularly interested in the small $\epsilon$ situation, in order to filter out eventual non-Markovian effects that can be linked to the finite size of the reservoir, the analysis here presented based on divisibility is more general and apply for any physical value of $\epsilon_i$ {\color{black} (still bounded by $0\leq \epsilon_i \leq 1$ and $0\leq \sum_i \epsilon_i \leq 1$)}. The difference is that for large values of $\epsilon$ we can no longer talk about a time differential equation for $\rho$ but rather about the divisibility of a discrete set of maps. {\color{black} Furthermore, it is possible to show that the dominating term of the eigenvalue $\lambda$ that determines the non-Markovian behavior is $-8aQ\epsilon_1\epsilon_2$. This means that the non-Markovian effect will always increase with $\epsilon$. Note that in our model, $\epsilon_i = \gamma_i \Delta t$ which means that larger values of $\epsilon_i$ can be understood either as larger decoherence rates and/or larger time intervals, both situations in which non-Markovian effects are known to exist.}

\textit{Non-Markovianity and Backflow of Information.} A question that often arises when studying non-Markovianity is that of the backflow of information from the environment to the system. Quantities like the distinguishability of two arbitrary states, among others, can witness such behavior and, therefore, are constantly linked to these problems. Our model is a simple and easily computable example of non-Markovian evolution that cannot be witnessed by these quantities because, in fact, there is no backflow of information in it. A quick look at Eq.~(\ref{rho3}) shows that the effect of the uncorrelated channels produces a contractive map of the order of $\epsilon$ in each time step, i.e. it typically deforms the Bloch Sphere that represents the possible initial states of the system into an ellipsoid whose axes shrink proportionally to $\epsilon$ after each collision. Since the effects of correlation that generate non-Markovianity are of the order of $\epsilon^2$, any quantity based on distances in the parameter space (Bloch sphere) will not witness any non-Markovian behavior in our model. {\color{black} Another way of saying this is that the bona-fide measure of information for a qubit is its von Neumann entropy $S=-\textrm{Tr} (\rho \log \rho)$. In our results, $S$ always increases after each collision, i.e. the information is always ``flowing'' from the system to the reservoir. This can be easily explained by the fact that the eventual non-Markovian effect, related to a particular negative $\gamma_i$ rate is always of the order of $\epsilon ^2$ while the standard Markovian channels are of the order of $\epsilon$. Therefore, even though the evolution may not be given by a CP map (may not be Markovian), the map is always contractive leading to guaranteed loss of information on the qubit after each collision. Furthermore, as we have shown, correlations in the environment change the rate of flow of the information but these changes happen both for the Markovian and non-Markovian situations.} 

Recent works have also related non-Markovianity and backflow of information to the entanglement (or discord) created between the system and the environment as, for example, in~\cite{rivas, alipour}. This can also be easily analyzed in our model because we can assume, with no loss of generality, initially uncorrelated pure states for system and environment. In this case the correlation between them due to the time evolution is simply the entanglement $E(t)$ created by the collisions which is given by the von Neumann entropy of the system $S=-\textrm{Tr}(\rho \log \rho)$. Under these conditions, the $\epsilon$ contraction described in the last paragraph also means that, in general, the system is getting more entangled with the environment in each time step, the exceptional cases being those in which the state of the system simply does not evolve in time like, for example, when there is only one channel $\sigma$ and the initial state is an eigenstate of this operator. These exceptions are trivial Markovian examples, though, and not interesting for us. First of all, note that this ever increasing entanglement means that our model provides another counter-example, one that can be easily tested in the labs by the way, to the results suggested in~\cite{alipour} and~\cite{shabani} and already disproved in~\cite{brodutch} relating Markovian evolution to vanishing discord between system and reservoir. Because our initial state is pure and the overall evolution unitary, the discord between both systems is equal to the entanglement which always increase in each time step. 
\begin{figure}[htb]
	\centering
		\includegraphics[width=0.35\textwidth]{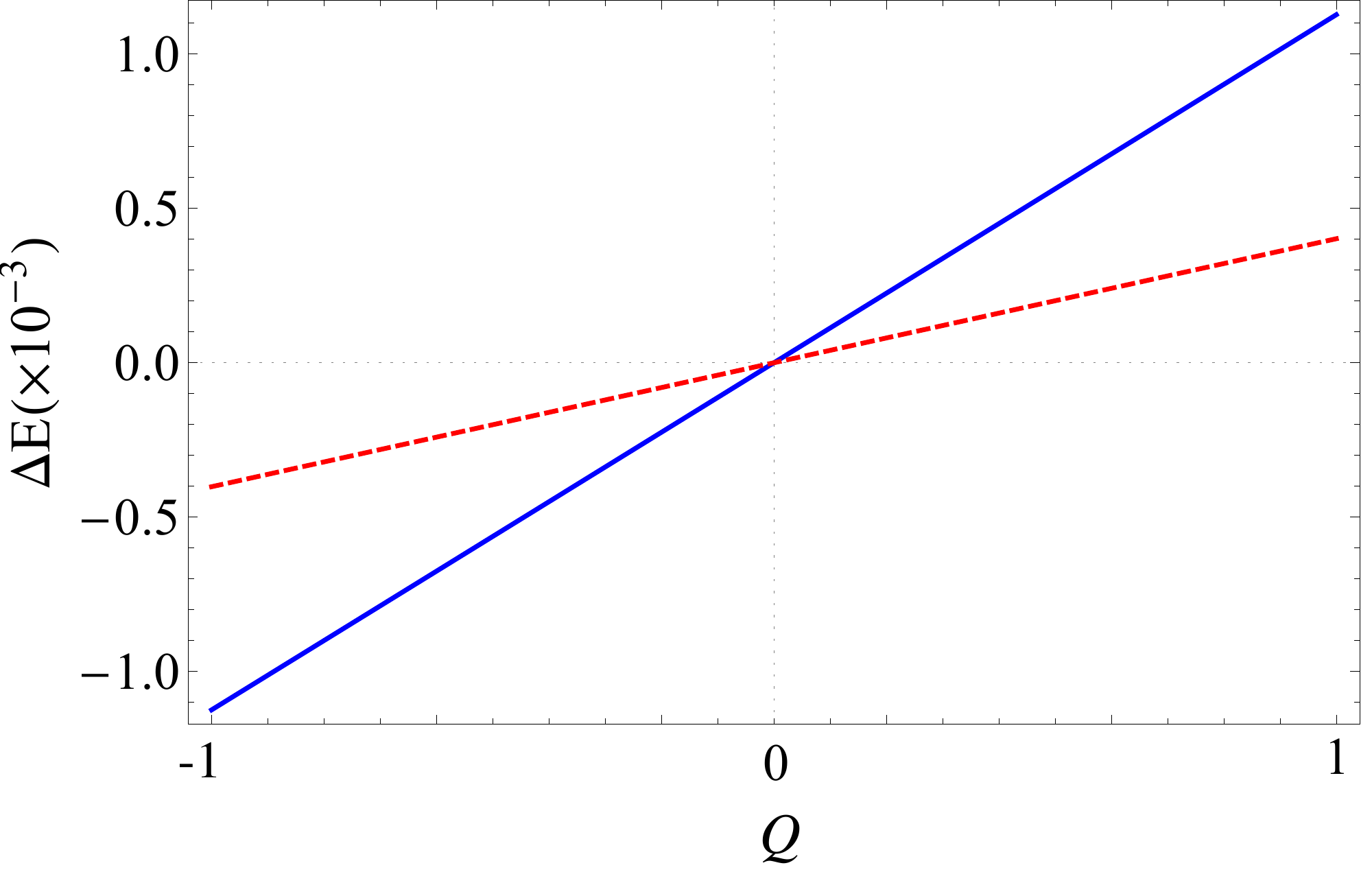}
	\caption{(Color online) The difference of entanglement $\Delta E$ versus the correlation factor $Q$ for $a=1$ and $\epsilon_1=\epsilon_2=0.01$ with initial states belonging to the $xz$-plane of Bloch sphere (blue solid line) and for $a=0$ and $\epsilon_1=0.01$ and $\epsilon_2=0.02$ with initial states with some component in the $z$-direction (red dashed line).}
	\label{D}
\end{figure}

There are, however, corrections of the contraction rate, proportional to $\epsilon^2$, that are due to the correlation factor of the environmental state. These corrections can be made explicit by calculating the difference $\Delta E = E(2\Delta t)-E_{Q=0}(2\Delta t)$ between the real entanglement of the system and environment for the second time step and that of an uncorrelated evolution ($q=1/2$). In Fig.~(\ref{D}) we display this difference for the two meaningful examples analyzed in the text: $\{a=0,\epsilon_1 \neq \epsilon_2\}$ and $\{a=1,\epsilon_1 = \epsilon_2\}$. In each case we choose the initial state that accentuate this difference the most and the figure for all possible initial states is in the Appendix. Note that in both cases, the correlation factor defines the sign of the correction to the entanglement rate for the second time step, but because  the evolution is still Markovian for $a=0$, this cannot be used to establish any direct relation between this rate and the character of the dynamics of the system. In this sense, our model also provides an example of non-Markovian behavior that cannot be detected by the entanglement between the system and the environment.

\begin{figure}[htb]
	\centering
		\includegraphics[width=0.35\textwidth]{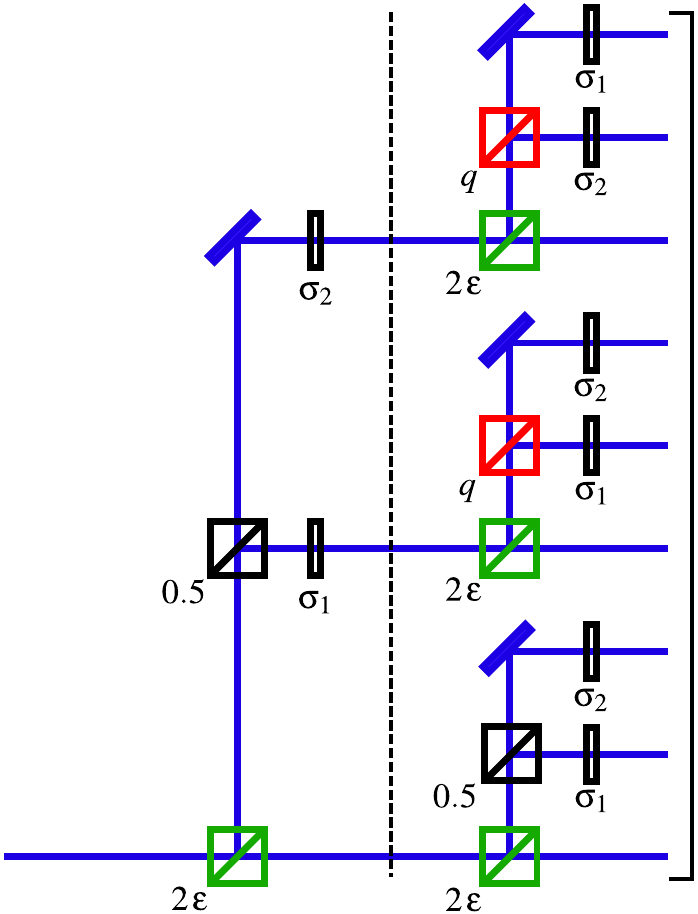}
	\caption{(Color online) Simple quantum optical experimental setup for the implementation of the collisional model. The blue lines represent the possible paths for the polarized photon. The photon passes through the setup where beam splitters, represented by squares, establish the probability of the action of the half-wave plate (black rectangles). The beam splitters have reflectivity of $2\epsilon$ (green color), of 0.5 (black color), or $q$ (red color). The half-wave plates can have their optical axis at $(\arcsin{\sqrt{a}})/2$ with respect to the $x$ direction ($\sigma_1$) or at $0^{\circ}$ ($\sigma_2$). The dashed black line is just for illustrative reasons and it delimits the first collision. After passing through this setup, the system's state should be tomographically determined, which is not represented in this figure.}
	\label{figexp}
\end{figure}

{\color{black} \textit{A simple quantum optical example.} Before concluding, we would like to present a simple quantum optical setup where this kind of analysis can be implemented. Assume a source of single photons, either on demand or heralded (such as parametric down conversion). The setup shown in Fig.~\ref{figexp} implements exactly the desired time evolution for the polarization qubit of the incoming photon. For simplicity, we show just the case where $\epsilon_1=\epsilon_2=\epsilon$, but the most general case can be realized with a similar scheme. The environment state is encoded in the different possible paths that the photon may follow. The operation $\sigma_1$ is given by a half-wave plate with the fast axis at an angle of $(\arcsin{\sqrt{a}})/2$ with respect to the $x$ direction and the operation $\sigma_2$ is given by a half-wave plate with the fast axis at $0^{\circ}$. The reflectivities of different beam splitters (indicated at their lower left corners) establish the probabilities that the system will suffer a change. The first collision is represented in the setup with the operations before the dashed (black) line: with a probability of $1-2\epsilon$ the photon will not suffer any change, but with probability $\epsilon$, its polarization will suffer one of the possible rotations ($\sigma_1$ or $\sigma_2$). For the second collision, the scheme follows the same logic. In order to measure the system's state, the outputs of the setup should be combined in a detector and quantum state tomography should be realized. By reconstructing the state after one and two collisions, we can verify how the dynamics of evolution is.}  

\textit{Conclusions.} In this work, we have designed a new collisional model that provides an intuitive way to approximate the mathematical definition of Markovian evolution into a feasible quantum optical experiment. For this model, the smallest set of requirements to simulate non-Markovian dynamics has been established. We have also studied the effects of correlations in the environmental states in the dynamics of the system and, in particular, we have shown that correlations alone are not sufficient to generate non-Markovianity, which will also depend on the particularities of the interaction between system and reservoir, and we have analyzed under which conditions this happens in our model. In order to do so, we have derived the map that describes the evolution of the system and checked its complete positivity as a function of the parameters of the model. We have also shown an example of non-Markovian evolution that does not violate the criterion of Ref.~\cite{breuer2}. Our case also exemplified that Markovian dynamics appear in system-environment states that are not only correlated (with non-vanishing discord) but, in fact, whose correlation increases with time, contradicting Ref.~\cite{alipour, shabani}. Furthermore, we have showed an example where the correlation in the environment modulates the rate at which it entangles with the system but this modulation is detached from the particular character of the dynamics and, therefore, cannot be used as a witness of non-Markovianity. Finally, we would like to stress that the model we have designed is very simple and yet physically meaningful: for example, with some adaptations it could be used to study situations like diluted gases where a single particle collides consecutively with others, one at a time. This can be a possible line of extension of the current work.

\subsection*{Acknowledgments}
The authors would like to thank P. Haikka for useful discussions on non-Markovianity. N.K.B, C.H.M. and M.F.S. would like to thank the support from the Brazilian agencies CNPq and CAPES. M.F.S. would like to thank the support of FAPEMIG, project PPM IV. This work is part of the INCT-IQ from CNPq and also of the Australian Research Council Centre of Excellence for Quantum Computation and Communication Technology (project number CE110001027).

\appendix

\section*{Appendix A}\label{app.A}

The action of a quantum map on a state can be characterized by $\Phi(\rho)=\sum_i\lambda_iT_i\rho T_i^{\dagger}$ where $\{\lambda_i\}$ are the eigenvalues of the matrix $\mathcal{H}$ \cite{aiello}. The operators $\{T_i\}$ are formed by the eigenvectors $\{u_i\}$ of $\mathcal{H}$ as follows
\be
T_i=
\begin{pmatrix}
\left[ \vec{u}_i\right]_0 & \left[ \vec{u}_i\right]_1\\
\left[ \vec{u}_i\right]_2 & \left[ \vec{u}_i\right]_3
\end{pmatrix},
\ee
where $\left[ \vec{u}_i\right]_j$ is the $j$th element of the vector $\vec{u}_i$. For a CP map $\{\sqrt{\lambda_i}T_i\}$ are the Kraus operators.

We would like to calculate the map that describes the evolution of the system from the first to the second collision for $\epsilon_1=\epsilon_2=\epsilon$. It is possible to show that
\ba
&&\rho(2\Delta t)=(1-2\epsilon)\rho(\Delta t)+C_1\left[\rho(\Delta t)-\sigma_y\rho(\Delta t)\sigma_y\right]+\nonumber\\
&&C_2\sigma_z\rho(\Delta t)\sigma_z+C_3\sigma_x\rho(\Delta t)\sigma_x+\\
&&C_4\left[\sigma_x\rho(\Delta t)\sigma_z+\sigma_z\rho(\Delta t)\sigma_x\right],\nonumber
\ea
where $\{C_i\}$ are functions of the parameters $a$, $\epsilon$ and $q$ and are given by
\ba
C_1=\frac{2a(2q-1)\epsilon^2(1-2\epsilon)}{1+4\epsilon(a\epsilon-1)},\nonumber\\
C_2=\frac{\epsilon  \{2 +8 \epsilon a- 32 a \epsilon^2 [\epsilon [(2 a -1)(q-1)- q+2]+1]\}}{1+4 \epsilon  (a \epsilon -1)}\nonumber\\
C_3=\frac{a \epsilon  \{1-4 \epsilon -4 \epsilon^2 [(2 a -1)(q-1)-q]\}}{1+4 \epsilon  (a \epsilon -1)},\nonumber\\
C_4=\frac{\sqrt{(1-a) a} \epsilon  \{1-4 \epsilon  [2 a (q-1) \epsilon +1]\}}{1+4 \epsilon  (a \epsilon -1)}.\nonumber
\ea

\begin{figure}[t]
			\includegraphics[width=0.35\textwidth]{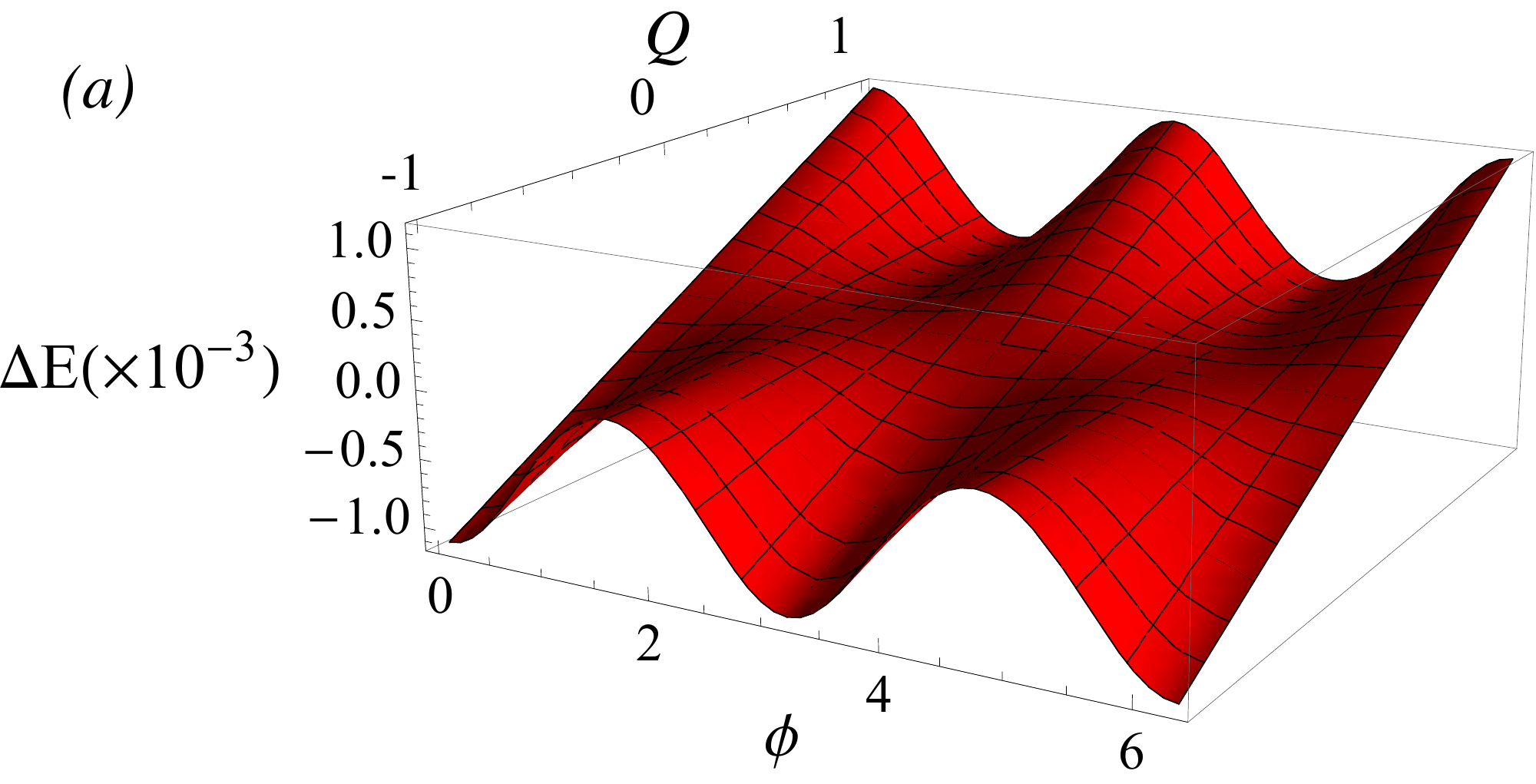} \includegraphics[width=0.35\textwidth]{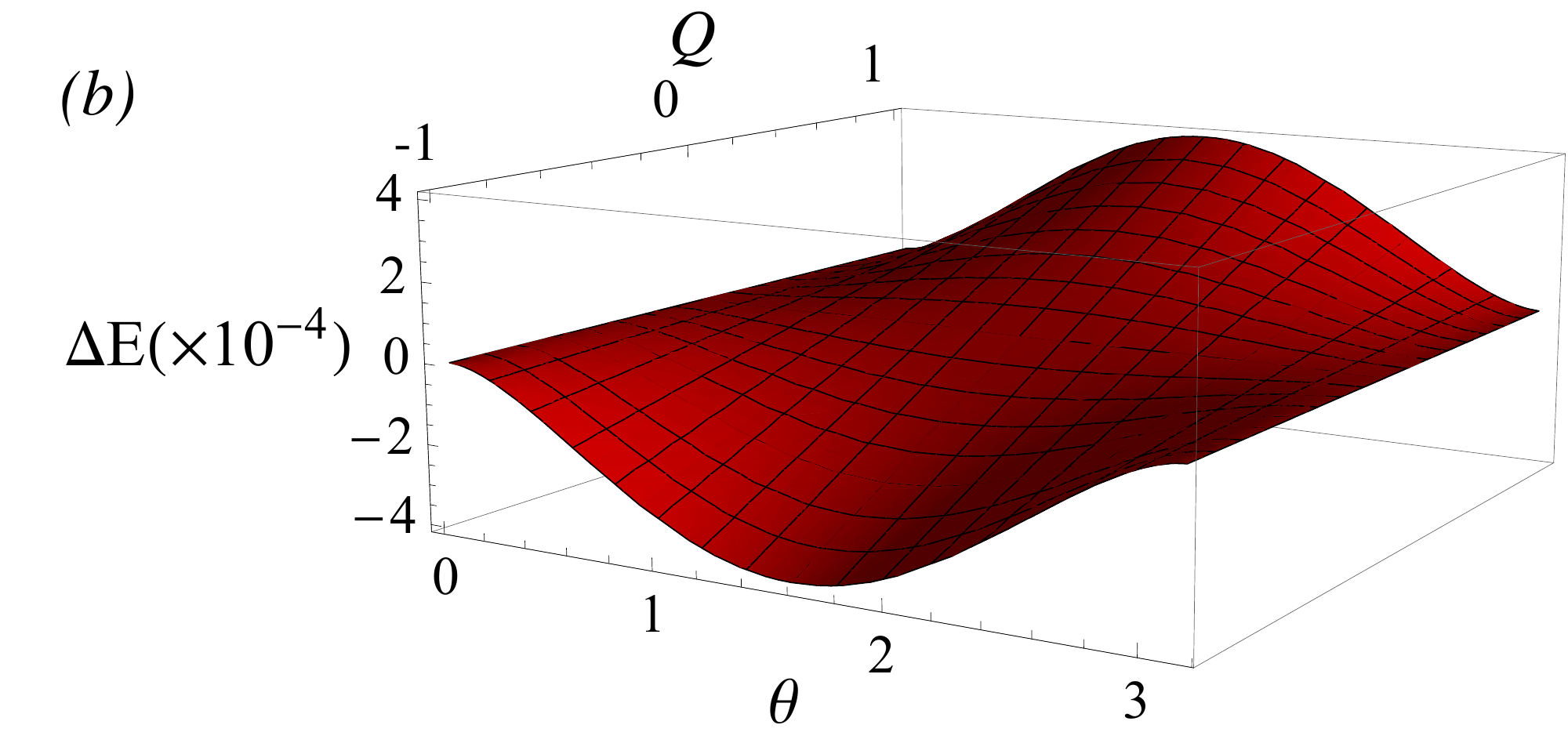}
	\caption{(Color online) The difference $\Delta E$ versus the correlation factor $Q$ for (a) $a=1$ and $\epsilon_1=\epsilon_2=0.01$ with initial states belonging to the $xz$-plane of Bloch sphere and (b) for $a=0$ and $\epsilon_1=0.01$ and $\epsilon_2=0.02$ with initial states with some component in the $z$-direction. The Bloch vector is defined as $\vec{r}=(\sin{\theta}\cos{\phi},\sin{\theta}\sin{\phi},\cos{\theta})$.}
	\label{discord}
\end{figure}

Notice here that because $\epsilon \ll 1$ and $0 \leq a \leq 1$, the only coefficient whose sign depends on the correlation factor $Q$ is $C_1$. All the other coefficients will be either zero or positive numbers. When $Q<0$ ($q<1/2$), $C_1$ is negative and the $\sigma_y$ term in Eq. (9) will be in the Lindblad form. However, when $Q>0$ ($q>1/2$), $C_1$ is positive and the evolution of the system cannot be described by a master equation in the Lindblad form anymore. This becomes clearer if we expand the coefficients until terms of the second order of $\epsilon$, assuming $\epsilon$ sufficiently small, as follows
\begin{widetext}
\ba
\rho(2\Delta t)\approx(1-2\epsilon)\rho(\Delta t)+
(2 - a) \epsilon\sigma_z\rho(\Delta t)\sigma_z+a\epsilon\sigma_x\rho(\Delta t)\sigma_x+\sqrt{a(1-a)}\epsilon\left[\sigma_x\rho(\Delta t)\sigma_z+\sigma_z\rho(\Delta t)\sigma_x\right]+\\
2aQ\epsilon^2\left[\rho(\Delta t)-\sigma_y\rho(\Delta t)\sigma_y\right]
=\nonumber\\
(1-2\epsilon)\rho(\Delta t)+
\epsilon\sigma_1\rho(\Delta t)\sigma_1+\epsilon \sigma_2\rho(\Delta t)\sigma_2+2aQ\epsilon^2\left[\rho(\Delta t)-\sigma_y\rho(\Delta t)\sigma_y\right].\nonumber
\ea
\end{widetext}

\section*{Appendix B}\label{app.B}

Fig.~\ref{discord} shows the difference $\Delta E = E(2\Delta t)-E_{Q=0}(2\Delta t)$ between the entanglement of the system and the environment in the actual evolution (as a function of $Q$) and that obtained when $Q=0$ and as a function of different initial states of the system, defined by Bloch vectors given by $\vec{r}=(\sin{\theta}\cos{\phi},\sin{\theta}\sin{\phi},\cos{\theta})$. Either in the case $\epsilon_1=\epsilon_2$ and $a=1$ or in the case $\epsilon_1\neq\epsilon_2$ and $a=0$, the environment where its qutrits are more correlated ($Q>0$) will result in a system less correlated to its environment than in the case of the equivalent uncorrelated evolution ($Q=0$). On the other hand, for $Q<0$, the system will become even more correlated to its environment than for $Q=0$ after the second collision. The entanglement between system and environment shows a similar behavior in both cases, but only in the first case ($\epsilon_1=\epsilon_2$, $a=1$) there is a Markovian to non-Markovian transition in the evolution of the system, therefore, the analysis of $\Delta E$ alone is not enough to characterize the dynamics of the system.


\end{document}